# The Distance to IC 10 from Near-Infrared Observations of Cepheids

Christine D. Wilson[1,2], Douglas L. Welch[1,2], I. Neill Reid[3,2] Abi Saha[4], and John Hoessel[5]

[1]Department of Physics and Astronomy, McMaster University, Hamilton, Ontario, Canada L8S 4M1

[2]Visiting Astronomer at the Canada-France Hawaii Telescope, operated by NRC Canada, CNRS France, and the University of Hawaii

[3]California Institute of Technology

[4]Space Telescope Science Institute

[5]University of Wisconsin, Madison

## ABSTRACT

We have measured the distance to the Local Group dwarf irregular galaxy IC 10 from near-infrared $JHK$ observations of Cepheids. The reddening-corrected distance modulus is $24.57 \pm 0.21$ mag or $0.82 \pm 0.08$ Mpc. This distance lies at the short end of the range of previous estimates, and is significantly more accurate than all prior determinations. At this distance, IC 10 is comparable in total mass to the Small Magellanic Cloud, but has much more vigorous star formation.

## 1. Introduction

The nearby galaxy IC 10 is important for studies of the interstellar medium and star formation in dwarf irregular galaxies, because its unusually strong and blue-shifted CO lines (Becker 1990) make it relatively easy to trace the distribution of both molecular and atomic hydrogen in this galaxy. However, studies of this galaxy are hindered by the fact that its distance is very poorly determined. Distance estimates range from a high of 3 Mpc based on the size of its HII regions (Sandage & Tammann 1974) to more commonly adopted values of 1-1.5 Mpc, which are estimated from the large HI extent of the galaxy (Roberts 1962), the degree of resolution into stars (Yahil, Sandage, & Tammann 1977), and, most recently, the luminosity of the Wolf-Rayet stars (Massey & Armandroff 1995). Recently, Wilson (1995) obtained a distance of 240 kpc based on near-infrared observations of candidate Cepheid variables. The main difficulty in determining the distance to this galaxy is that it lies extremely close to the Galactic plane (latitude = -3°), and thus has a high and possibly variable foreground reddening.

One way to reduce the effect of foreground reddening is to work in the near-infrared. In particular, Cepheid variable stars have nearly constant $H-K$ color with both phase and period (Welch et al. 1984), which makes it possible to determine the reddening from two-color single phase observations and so obtain the true distance to the galaxy. In this letter, we present $JHK$ observations for the four Cepheid variables in IC 10 identified recently at Gunn $r$ band by Saha, Hoessel, & Krist (1995). The $J-H$ colors show that some of the variables used by Wilson (1995) are not Cepheids but rather more luminous, cooler stars, and thus the distance to IC 10 is not as small as the preliminary estimate indicated.

## 2. Observations

Four Cepheids and five other variables in IC 10 identified by Saha et al. (1995) were observed at the Canada-France-Hawaii Telescope using the REDEYE infrared camera. The camera was used in narrow-field mode, which gives a scale of 0.2″ pixel$^{-1}$ and a 51″ field of view. One Cepheid was observed in $H$ and $K$ only on 1993 August 29 as part of a five-night run. Due to a technical problem with the narrow field dewar, the narrow-field optics were used to feed the wide-field dewar, which contains a cosmetically cleaner Hg:Cd:Te array. The remaining three Cepheids were observed in $JHK$ on 1994 October 25. Details of the observations are given in Table 1. Flat fields were obtained from images taken in an outer field of M31 during each run. There were small variations of unknown origin in the flat fields from night to night, and so the flat fielding is estimated to be accurate to only 1-2%.

Standard stars from the UKIRT Faint $JHK$ Standards list were observed during both runs. Three nights of the 1993 run (including 1993 August 29) were photometric. The zeropoints of the photometry were determined for each night separately from observations of four to six standard stars and are formally accurate to $\pm0.011$ to $\pm0.026$ mag. The maximum change in the zeropoint from one night to the next was 0.02 mag at $H$ and 0.11 mag at $K$. Eight standards were observed on 1994 October 25 and the night was photometric, with the zeropoint of the photometry formally accurate to $\pm0.010$ to $\pm0.015$ mag. For both runs, no color terms were found and we adopted extinction coefficients of 0.05 mag/airmass at $H$ and 0.10 mag/airmass at $K$. IC 10 was observed between 1.3 and 1.5 airmasses, while the standard stars were observed between 1.0 and 2.0 airmasses.

Photometry for all the stars on the frames was obtained using the packages DAOPHOT II, ALLSTAR, and ALLFRAME. Aperture corrections to correct for emission beyond our 4-6 pixel fitting radius were calculated from aperture photometry with a 20 pixel radius aperture for 1-4 bright stars in each field, and using an image from which all other stars had been removed. The aperture corrections are accurate to $\pm0.007$ to $\pm0.054$ mag. The identifications of the Cepheids on the infrared images were confirmed by cross-identifying several stars on both the infrared image and one of the $r$ images used in the initial optical identification and calculating the co-ordinate transformation between the optical and infrared images.

Five additional variables were observed on photometric nights during the two runs. The photometry for these stars is given in Table 2. From their location on the infrared color-color diagram (Figure 1), these objects are not Cepheids but are redder, more luminous stars, perhaps long period variables. Indeed, the light curves of these objects can be fit with periods of order one year, as well as with convincing Cepheid light curves (Saha et al. 1995). The $K$ luminosities and optical colors of these stars are similar to the two bright stars used by Wilson (1995) to obtain the very small distance to IC 10. Thus the error in that analysis was that the objects used to determine the distance were not in fact Cepheids.

## 3. Analysis

The calibrated $JHK$ magnitudes for the four Cepheids are given in Table 2. The uncertainties in Table 2 were obtained by summing in quadrature the uncertainties from the point spread function photometry, the aperture correction, the calibration zeropoint, and the flat fielding. The periods, absolute magnitudes, color excess, reddening, and true distance modulus are given in Table 3. The absolute magnitudes were calculated from

$$M_J = -2.15 - 3.294 \log_{10}(P)$$

$$M_H = -2.36 - 3.319 \log_{10}(P)$$

$$M_K = -2.38 - 3.374 \log_{10}(P)$$

obtained from Cepheids in the Magellanic Clouds (Welch et al. 1987, distance modulus $18.57 \pm 0.05$ mag), where $P$ is the period in days. Since the near-infrared colors have large uncertainties due to the faint magnitudes of the Cepheids, we chose to compute the infrared reddenings using $E(<r> -K)$. We take the $<r>$ magnitudes from Saha et al. (1995) and adopt

$$M_r = -2.91(log_{10}P - 1) - 4.04$$

(Hoessel et al. 1994). The reddenings and total absorptions were estimated using the absorption curve parameterization of Cardelli, Clayton, & Mathis (1989). Since the extinction law is not identical along all lines of sight, even in our own Galaxy, it is desirable to select a parameter which can be used to scale the law appropriately. Cardelli et al. found that $R_V = A_V/E_{B-V}$ worked well for this purpose. We assume, for want of better information, that the $R_V$ appropriate for the dust along the line of sight to IC 10 lies in the range 3.0–3.4. Over this range, we obtain the relation

$$A_K = 0.16E(<r> -K)$$

where, for comparison, $E(B - V) = (0.42 \pm 0.03)E(<r> -K)$.

The reddening-corrected $K$-band period-luminosity relation for IC 10 is shown in Figure 2. We obtain the distance to IC 10 by calculating the weighted average of the distances for the four Cepheids using the reddenings obtained from $E(<r> -K)$. We do not make any correction for phase. This method gives a true distance modulus of $24.57 \pm 0.21$ mag, or $0.82 \pm 0.08$ Mpc. The foreground reddening toward the Cepheids ranges from $E(B - V) = 0.6$ to 1.1 mag with an average value of 0.8 mag, in good agreement with previous studies.

There have been recent hints that the distance to IC 10 must be closer to 1 Mpc than 3 Mpc. In their study of the Wolf-Rayet stars, Massey, Armandroff, & Conti (1992) noted that the brightest Wolf-Rayet stars were roughly two magnitudes more luminous than the brightest Wolf-Rayet stars in the Large Magellanic Cloud (for an adopted distance modulus of 25.5 mag (1.25 Mpc) and a reddening of $E(B-V) = 0.87$ mag). They speculated that the galaxy was actually at a distance of 1.5 Mpc and had a much lower reddening than normally adopted. Massey & Armandroff (1995) used $BV$ photometry to estimate individual reddenings for the three brightest WC stars and obtain a distance estimate of $\sim 1$ Mpc by comparing their average luminosity to that of the brightest WC stars in the Large Magellanic Cloud. Adopting a foreground reddening of $E(B-V) = 0.87$ mag, Karachentsev & Tikhonov (1993) obtained a distance of 1 Mpc from photometry of the brightest red supergiants. Finally, the large negative radial velocity of IC 10 ($\sim -330$ km s$^{-1}$) suggests that IC 10 is strongly bound to the Local Group and has a trajectory with a large line-of-sight component.

## 4. Implications of the Distance to IC 10

Our new distance to IC 10 results in a dynamical mass in the central HI disk of $\sim 6 \times 10^8$ M$_\odot$ (comparable to that of the Small Magellanic Cloud), an HI mass of $5 \times 10^7$ M$_\odot$ (Shostak & Skillman 1989), and a far infrared luminosity of $\sim 10^8$ L$_\odot$ (Thronson et al. 1990). Since IC 10 is a low-metallicity system ($12 + log(O/H) = 8.04$, Lequeux et al. 1979) but has strong CO lines (Becker 1990), it is an important system for determining the effect of changes in metallicity on the conversion factor between CO flux and $H_2$ column density. With the new distance to IC 10, the CO-to-$H_2$ conversion factor in IC 10 is 2.5 times larger than the Galactic value (Wilson & Reid 1991). Thus the data for IC 10 are consistent with the analysis of recent data for the Small Magellanic Cloud (Rubio, Lequeux, & Boulanger 1993), which implies the CO-to-$H_2$ conversion factor is larger and hence that the CO lines are weaker for a given amount of molecular gas in low-metallicity systems. Since IC 10 has not been completely mapped in CO, it is difficult to estimate the total molecular gas mass.

However, adopting a CO-to-$H_2$ conversion factor 2.5 times that of our own Galaxy, the molecular gas mass contained in the peaks observed by Becker (1990) is $M_{mol} = 1.4 \times 10^7$ $M_\odot$. For comparison, a complete map of the Small Magellanic Cloud yields a molecular gas mass of only $6 \times 10^6$ $M_\odot$ (Rubio et al. 1991), if a CO-to-$H_2$ conversion factor 4 times larger than the Galactic value is assumed (Rubio et al. 1993). Thus IC 10 is among the least massive and luminous of the Local Group dwarf irregular galaxies.

Despite its small size, the star formation properties of IC 10 are unusual. From radio continuum observations, Yang & Skillman (1993) identified an unusually large nonthermal superbubble in IC 10 that they suggested had been created by several supernovae. For a distance of 1.25 Mpc, this large superbubble had a diameter that was approximately four times larger than supernova remnants of comparable surface brightness in the Large Magellanic Cloud; at 0.8 Mpc it is still unusually large. If we correct the HII region luminosity function of Hodge & Lee (1990) to the new distance and reddening, the luminosity of the brightest HII region is $\sim 10^{38}$ erg, not particularly luminous compared to the giant HII regions in the LMC and M33 (Kennicutt 1988). However, IC 10 has an extremely high surface density of Wolf-Rayet stars (Massey et al. 1992), $\sim 7$ stars $kpc^{-2}$ at the new distance or somewhat higher than in two active fields in M33. Although IC 10 is a very small galaxy, it is still an extremely interesting one from the point of view of its star formation properties. Finally, comparing the distance to IC 10 with that to M31 ($0.71 \pm 0.03$ Mpc, Welch et al. 1986) suggests that the large negative radial velocity of IC 10 is perhaps due to its being bound to M31, and in the part of its orbit where its velocity is almost entirely along the line of sight. However, its distance from M31 in the plane of the sky is $\sim 250$ kpc.

## 5. Conclusions

We have measured the first accurate distance to the nearby dwarf irregular galaxy IC 10 from $JHK$ observations of four Cepheids. The Cepheids give a true distance modulus of $24.57 \pm 0.21$ mag for a distance of $0.82 \pm 0.08$ Mpc. The average (foreground + internal) reddening towards IC 10 obtained from the Cepheids is $E(B-V) \sim 0.8$ mag, close to the commonly adopted value of $E(B-V) = 0.87$ mag. At this distance, IC 10 has a total mass very similar to that of the Small Magellanic Cloud, but has much more vigorous star formation.

| Cepheids | Calendar Date | Filter | Julian Date (JD - 2440000) | T (s) | Airmass | FWHM (″) |
|---|---|---|---|---|---|---|
| V2 | 1994/10/25 | J | 9650.944 | 720 | 1.44 | 1.0 |
| | | H | 9650.953 | 540 | 1.48 | 1.0 |
| | | K | 9650.964 | 900 | 1.53 | 1.0 |
| V7,V8 | 1994/10/95 | J | 9650.904 | 720 | 1.34 | 0.9 |
| | | H | 9650.914 | 540 | 1.36 | 0.8 |
| | | K | 9650.924 | 900 | 1.39 | 0.8 |
| V10 | 1993/8/29 | H | 9228.993 | 1080 | 1.30 | 0.7 |
| | | K | 9229.011 | 1200 | 1.30 | 0.7 |

Table 1: Observing Log for Cepheids in IC 10

| Variable | $K$ (mag) | $H-K$ (mag) | $J-K$ (mag) | $J-H$ (mag) | $<r>$[1] (mag) |
|---|---|---|---|---|---|
| Cepheids: | | | | | |
| V2 | 18.97±0.15 | 0.39±0.21 | 0.70±0.16 | 0.31±0.16 | 22.25 |
| V7 | 17.89±0.07 | 0.03±0.08 | 0.58±0.07 | 0.55±0.06 | 21.23 |
| V8 | 17.43±0.07 | 0.21±0.08 | 0.92±0.07 | 0.71±0.06 | 21.05 |
| V10 | 17.43±0.08 | 0.48±0.09 | ... | ... | 21.89 |
| Others: | | | | | |
| V3 | 15.41±0.05 | 0.42±0.07 | 1.74±0.06 | 1.32±0.05 | 22.70 |
| V5 | 14.80±0.04 | 0.24±0.04 | 1.32±0.06 | 1.08±0.06 | 20.52 |
| V6 | 14.40±0.03 | 0.39±0.04 | 1.81±0.06 | 1.42±0.06 | 21.50 |
| V9 | 14.23±0.06 | 0.61±0.06 | ... | ... | 20.19 |
| V12 | 14.53±0.02 | 0.46±0.04 | ... | ... | 20.96 |

Table 2: Near-Infrared Magnitudes of Variables in IC 10

[1] From Saha et al. 1995.

| Cepheid | $\log_{10}(P)$ | $M_K$ (mag) | E(H-K) (mag) | $E(<r> - K)$ (mag) | $A_K$ (mag) | $(m-M)_o$ (mag) |
|---|---|---|---|---|---|---|
| V2 | 1.074 | -6.00 | 0.31±0.21 | 1.57±0.15 | 0.25±0.02 | 24.72±0.17 |
| V7 | 1.489 | -7.40 | -0.07±0.08 | 1.40±0.07 | 0.22±0.01 | 25.07±0.08 |
| V8 | 1.393 | -7.08 | 0.11±0.08 | 1.72±0.07 | 0.28±0.01 | 24.23±0.08 |
| V10 | 1.457 | -7.30 | 0.38±0.09 | 2.53±0.08 | 0.40±0.01 | 24.33±0.09 |

Table 3: Distance Modulus and Reddening to IC 10 for Individual Cepheids

This manuscript was prepared with the AAS LaTeX macros v3.0.

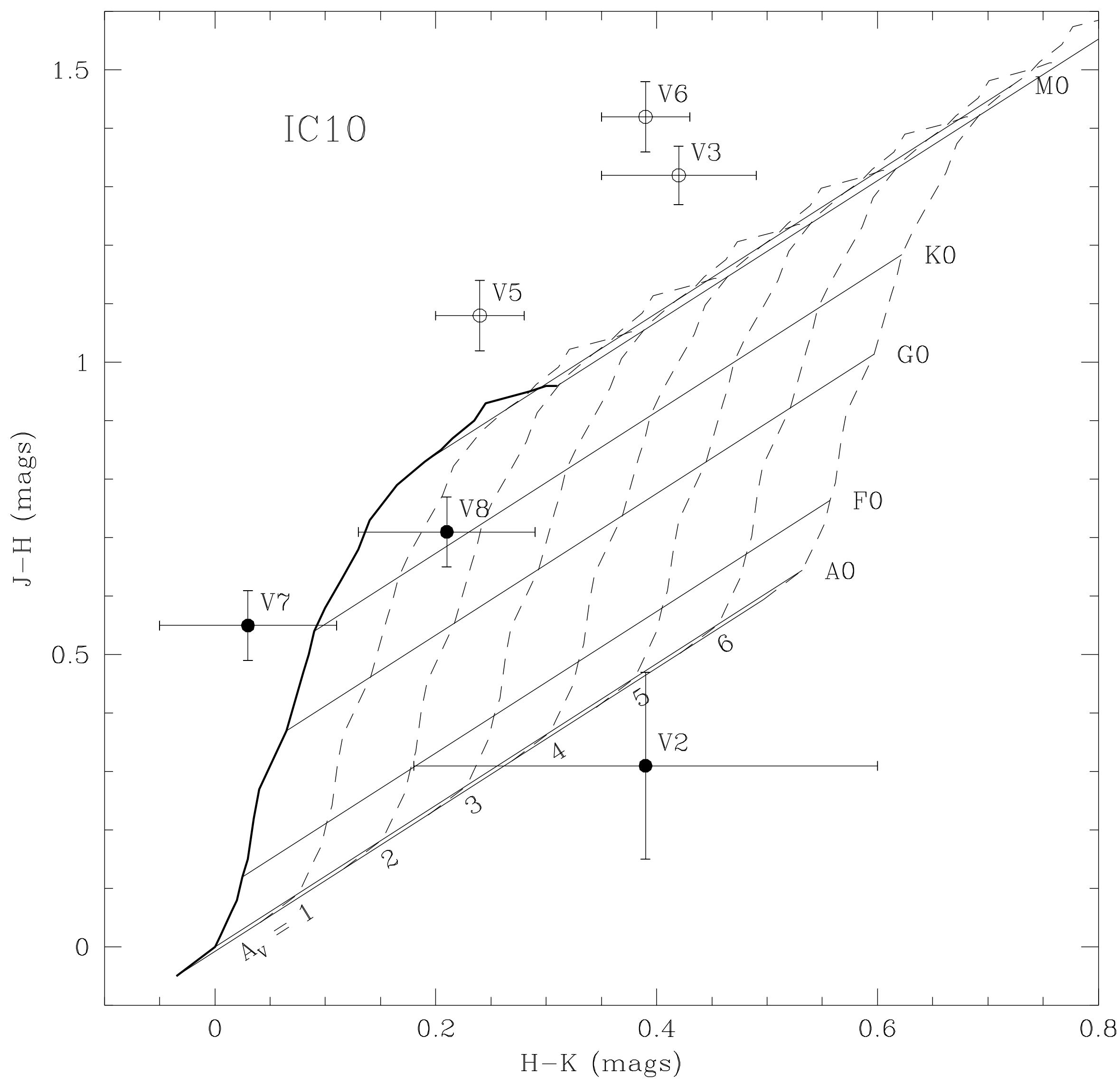

Fig. 1.— The infrared color-color plot for the variables in IC 10. The thin solid lines are lines of constant spectral type, while the dashed lines are lines of constant reddening. The heavy solid line indicates the unreddened main sequence. The Cepheids are plotted as solid symbols, while the other variables are plotted as open symbols.

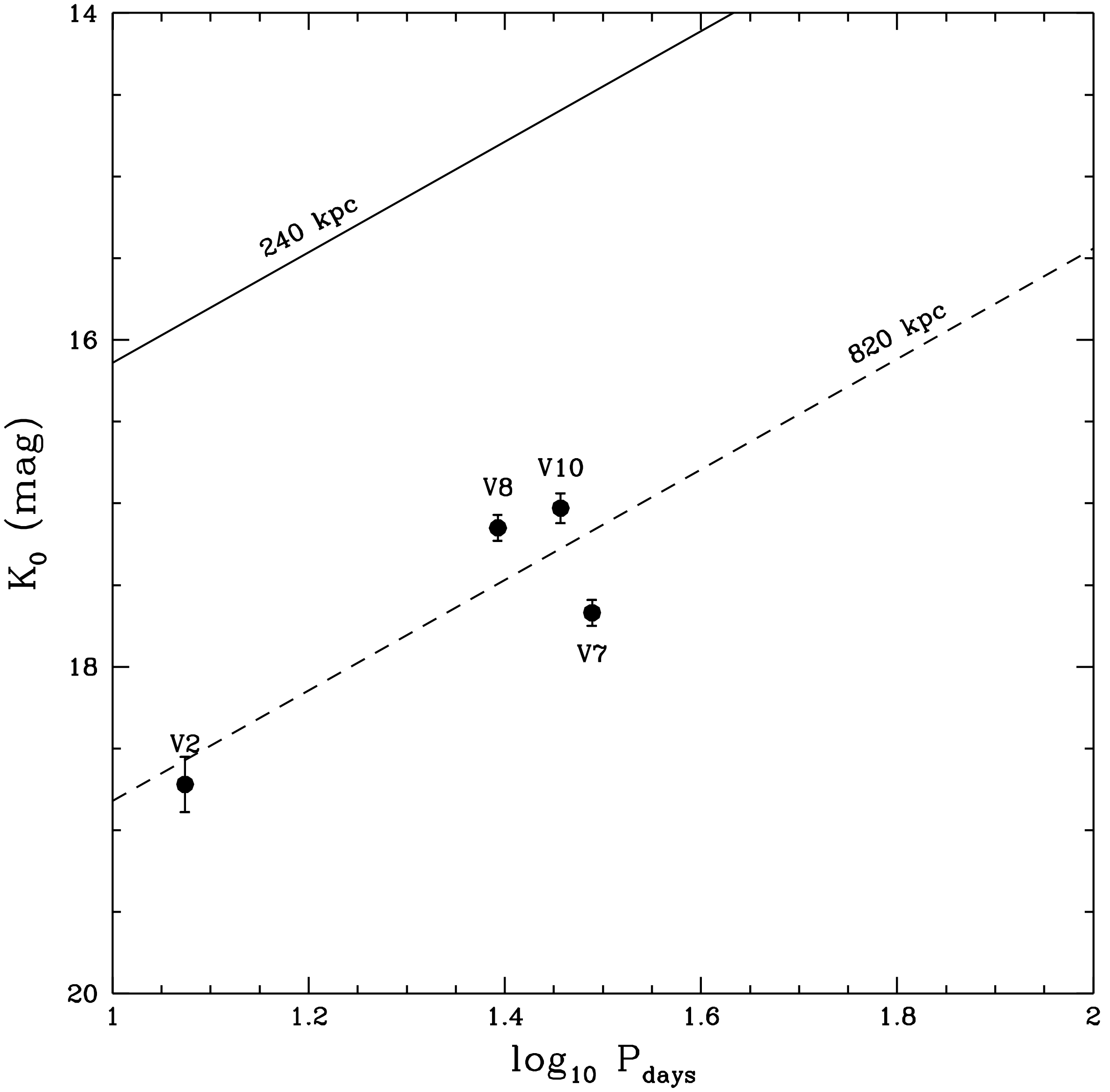

Fig. 2.— The reddening-corrected $K$-band period-luminosity diagram for the four Cepheids in IC 10. The dashed line indicates the period-luminosity relation if IC 10 is at a distance of 0.82 Mpc.